\documentclass[aps,prl,twocolumn,showpacs,amsmath,amssymb,superscriptaddress]{revtex4}
\usepackage{graphicx}
\usepackage{amssymb}
\usepackage{natbib}

\begin{document}
\title{Neutron spin resonance as a probe of superconducting gap anisotropy in partially detwinned
electron underdoped NaFe$_{0.985}$Co$_{0.015}$As}

\author{Chenglin Zhang}
\affiliation{Department of Physics and Astronomy, Rice University, Houston, Texas 77005, USA}

\author{J. T. Park}
\affiliation{Heinz Maier-Leibnitz Zentrum (MLZ),
Technische Universit\"{a}t M\"{u}nchen, D-85748 Garching, Germany}

\author{Xingye Lu}
\affiliation{Department of Physics and Astronomy, Rice University, Houston, Texas 77005, USA}
\affiliation{Institute of Physics, Chinese Academy of Sciences, Beijing
100190, China}

\author{Rong Yu}
\affiliation{Department of Physics and Beijing Key Laboratory of Opto-electronic Functional Materials and Micro-nano Devices, Renmin University of China, Beijing 100872, China}
\affiliation{Department of Physics and Astronomy, Collaborative Innovation Center of Advanced Microstructures, Shanghai Jiaotong University, Shanghai 200240, China}

\author{Yu Li}
\affiliation{Department of Physics and Astronomy, Rice University, Houston, Texas 77005, USA}

\author{Wenliang Zhang}
\affiliation{Institute of Physics, Chinese Academy of Sciences, Beijing
100190, China}

\author{Yang Zhao}
\affiliation{NIST Center for Neutron Research,
National Institute of Standards and Technology,
Gaithersburg, MD 20899-6102, USA}
\affiliation{Department of Materials Science and Engineering, University of Maryland, College Park, Maryland 20742, USA}

\author{J. W. Lynn}
\affiliation{NIST Center for Neutron Research,
National Institute of Standards and Technology,
Gaithersburg, MD 20899-6102, USA}

\author{Qimiao Si}
\affiliation{Department of Physics and Astronomy, Rice University, Houston, Texas 77005, USA}

\author{Pengcheng Dai}
\email{pdai@rice.edu}
\affiliation{Department of Physics and Astronomy, Rice University, Houston, Texas 77005, USA}

\begin{abstract}
We use inelastic neutron scattering (INS) to study the spin excitations in partially detwinned
NaFe$_{0.985}$Co$_{0.015}$As which has coexisting static antiferromagnetic (AF) order and superconductivity ($T_c=15$ K, $T_N=30$ K).
In previous INS work on a twinned sample, spin excitations form a dispersive sharp resonance near $E_{r1}=3.25$ meV and a broad
dispersionless mode at $E_{r1}=6$ meV at the AF ordering wave vector
${\bf Q}_{\rm AF}={\bf Q}_1=(1,0)$ and its twinned domain ${\bf Q}_2=(0,1)$. For partially detwinned NaFe$_{0.985}$Co$_{0.015}$As with
the static AF order mostly occurring at ${\bf Q}_{\rm AF}=(1,0)$,  we still find a double resonance
at both wave vectors with similar intensity.  Since ${\bf Q}_1=(1,0)$ characterizes the explicit breaking of the spin rotational symmetry associated with the AF order, these results indicate that the double resonance cannot be due to the static and fluctuating AF orders, but originate from the superconducting gap
anisotropy.
\end{abstract}


\maketitle

\begin{center}
\textbf{I. Introduction}
\end{center}

The neutron spin resonance is a collective magnetic excitation
observed by inelastic neutron scattering (INS)
at the antiferromagnetic (AF) ordering wave vector of
unconventional superconductors below $T_c$ \cite{mignod,Eschrig,jmtranquada,dai}. First
discovered in the optimally hole-doped YBa$_2$Cu$_3$O$_{6+x}$ family of copper oxide superconductors \cite{mignod},
the mode was also found in iron pnictide superconductors at the AF wave vector ${\bf Q}_{\rm AF}=(1,0)$ in
reciprocal space [Figs. 1(a) and 1(b)] \cite{kamihara,cruz,christianson,CZhang13,NQureshi14,lumsden,schi09,dsinosov09,steffens,MGKim13,CLZhang13,CLZhang14}, and is considered one of the hall marks
of unconventional superconductivity \cite{scalapino}. Experimentally, the neutron spin resonance appears
as an enhancement of the magnetic spectral weight at an energy $E_r$ in the superconducting
state at the expense of normal state spin excitations for energies below it.
For iron pnictide superconductors with hole and electron Fermi surfaces near the $\Gamma$ and $M$ points, respectively [Fig. 1(c)] \cite{hirschfeld,chubukov},
the mode is generally believed to arise from sign reversed
quasiparticle excitations between the hole and electron Fermi surfaces and occur at an energy
below the sum of their superconducting gap energies ($E_r\leq \Delta_h+\Delta_e$) \cite{Korshunov,Maier}.

If the energy of the resonance is associated with the superconducting gap energies at the hole and electron Fermi surfaces,
it should be sensitive to their anisotropy on the respective Fermi surfaces \cite{Maier09,Goswami}.  Indeed, recent
INS experiments on the NaFe$_{1-x}$Co$_x$As family of iron pnictide superconductors reveal the presence of a dispersive sharp resonance
near $E_{r1}=3.25$ meV and a broad dispersionless mode at $E_{r2}=6$ meV at ${\bf Q}_{\rm AF}=(1,0)$ in electron underdoped superconducting
NaFe$_{0.985}$Co$_{0.015}$As with static AF order ($T_c=15$ K and $T_N=30$ K) \cite{CLZhang13,CLZhang14}.
From the electronic phase diagram of NaFe$_{1-x}$Co$_x$As determined from specific heat \cite{GTTan2013}, scanning tunneling
microscopy \cite{PCai2013}, and nuclear magnetic resonance (NMR) \cite{SWOh2013,LMa2014} experiments, we know that 
NaFe$_{0.985}$Co$_{0.015}$As is a bulk superconductor with microscopically 
coexisting static AF ordered and superconducting phases.  
For Co-doping near 
optimal superconductivity around $x=0.0175$, NaFe$_{1-x}$Co$_x$As becomes mesoscopically phase separated with static AF ordered 
and paramagnetic superconducting phases \cite{LMa2014}, similar to the co-existing cluster  
spin glass and superconducting phases in optimally electron-doped BaFe$_2$As$_2$ \cite{Bernhard2012,XYLu2014B}. 
Since angle resolved photoemission spectroscopy (ARPES) experiments 
on NaFe$_{0.985}$Co$_{0.015}$As found a 
large superconducting gap anisotropy in the electron Fermi pockets \cite{qqge},
the double resonance may result from orbital-selective pairing induced superconducting gap anisotropy
along the electron Fermi surfaces \cite{RYu14}.  Upon increasing electron doping to $x=0.045$ to form
superconducting NaFe$_{0.955}$Co$_{0.045}$As ($T_c=20$ K), the superconducting gap anisotropy disappears \cite{Liu_arpes,thirupathaiah}
and INS reveals only a single sharp resonance coupled with superconductivity \cite{clzhang13b}.

Although the superconducting gap anisotropy provided a possible interpretation \cite{RYu14}, 
 the double resonance in underdoped
NaFe$_{0.985}$Co$_{0.015}$As may also be due to the
coexisting static AF order with superconductivity \cite{knolle11,WCLv14}.  Since
${\bf Q}_{\rm AF}={\bf Q}_1=(1,0)$ 
characterizes the explicit breaking of the spin rotational symmetry
in the AF ordered state of
a completely detwinned sample [Fig. 1(b) and 1(c)] \cite{YSong13}, one should expect magnetic susceptibility anisotropy
at ${\bf Q}_{\rm AF}={\bf Q}_1=(1,0)$ and ${\bf Q}_2=(0,1)$.  At the AF ordering wave vector ${\bf Q}_{\rm AF}={\bf Q}_1=(1,0)$,
the resonance appears in the longitudinal susceptibility, whereas the transverse
component displays a spin-wave Goldstone mode.  At the other momentum ${\bf Q}_2=(0,1)$, the resonance has both longitudinal and transverse
components and is isotropic in space.  If the resonance shows distinct energy scales at ${\bf Q}_1$ and ${\bf Q}_2$, one would expect to find a 
double resonance in a twinned sample as shown in Fig. 1(e) \cite{knolle11,WCLv14}.  However, one 
would then expect a single resonance of energy $E_{r1}$ at ${\bf Q}_1$ and that of energy 
$E_{r2}$ at ${\bf Q}_2$ in a completely detwinned superconducting sample with static AF order [Fig. 1(f)].

\begin{figure}[t] \includegraphics[scale=.4]{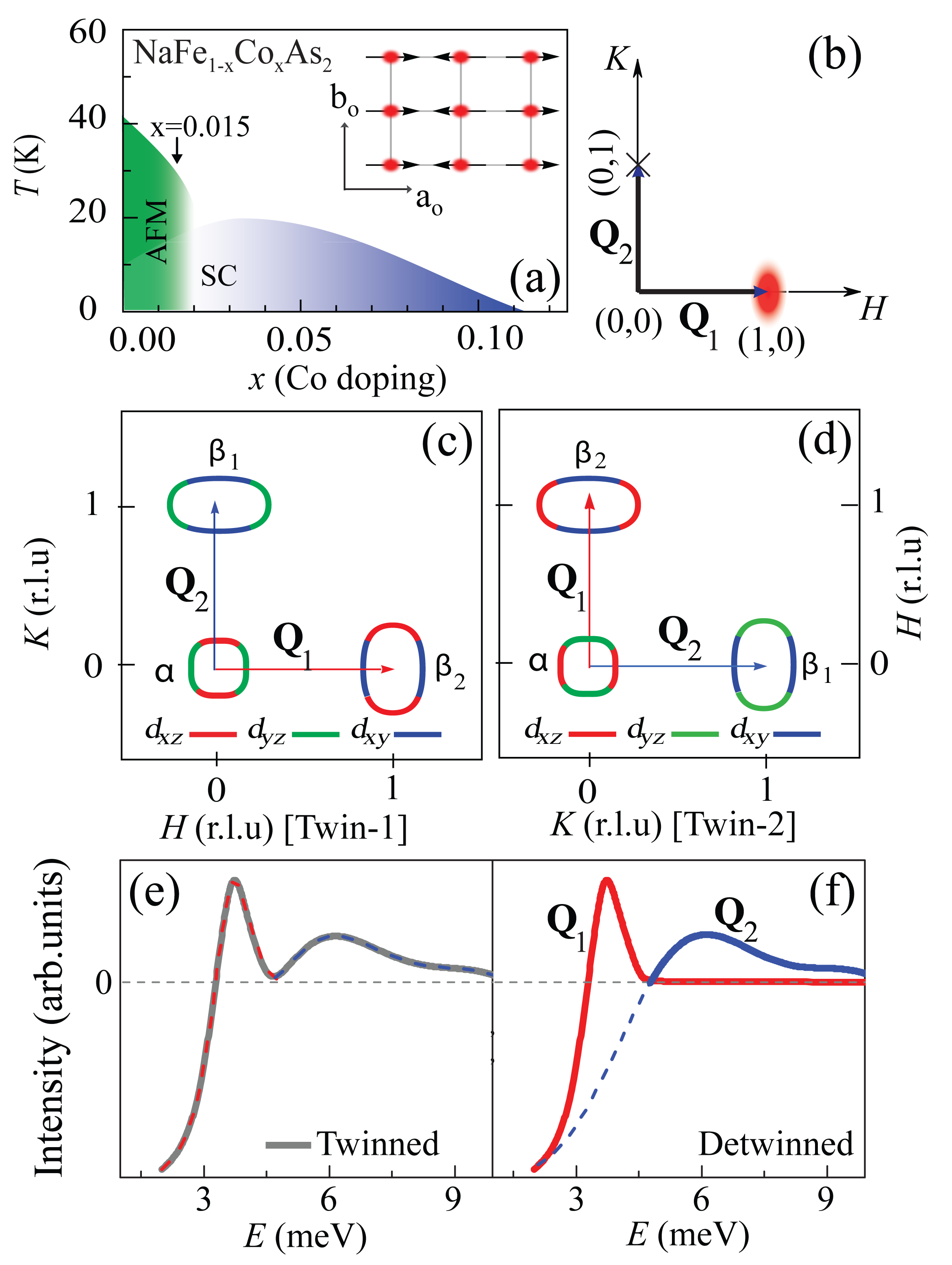}
\caption{(Color online)
(a) The electronic phase diagram of NaFe$_{1-x}$Co$_x$As, where the arrow indicates the Co-doping level of our sample. The inset shows the magnetic structure in orthorhombic notation. (b) In a 100\% detwinned sample, one should observe magnetic
order at ${\bf Q}_{\rm AF}={\bf Q}_1=(1,0)$ but not at ${\bf Q}_2=(0,1)$. (c) The schematic drawings of the
Fermi surfaces in a detwinned sample. (d) Its twin domain rotated 90$^\circ$ away.  The arrows mark
Fermi surface nesting wave vectors.  (e) In a twinned sample,
one cannot distinguish ${\bf Q}_1=(1,0)$ and ${\bf Q}_2=(0,1)$, and therefore there should
be two resonances at both wave vectors. (f) In the AF order and superconductivity coexisting theory,
$E_{r1}$ should appear at ${\bf Q}_1$ and $E_{r2}$ at ${\bf Q}_2$ in a completely
detwinned superconducting sample with static AF order at ${\bf Q}_1$.
}
\end{figure}

To test if this is indeed the case, we have carried out INS experiments on uniaxial strain partially 
detwinned NaFe$_{0.985}$Co$_{0.015}$As to study
the neutron spin resonance at ${\bf Q}_1$ and ${\bf Q}_2$.  Instead of $E_{r1}$ at ${\bf Q}_1$ and
$E_{r2}$ at ${\bf Q}_2$ as expected from the theory of coexisting static AF order with superconductivity \cite{knolle11,WCLv14}, we
find that both $E_{r1}$ and $E_{r2}$ are present at ${\bf Q}_1$ and ${\bf Q}_2$
as in the twinned case.  Therefore, the presence of the double resonance is not directly associated with the breaking of the spin rotational symmetry
in detwinned NaFe$_{0.985}$Co$_{0.015}$As.  Instead, our results are consistent with the notion that the splitting of the resonance is due to
superconducting gap anisotropy in the underdoped NaFe$_{0.985}$Co$_{0.015}$As, suggesting weak direct coupling between spin waves and
superconductivity.  These results are also consistent with polarized neutron scattering data, where
 the longitudinal spin excitations of $E_{r1}$ reveals a clear order-parameter-like increase below $T_c$ reminiscent of the resonance, while
the transverse spin excitations of the $E_{r1}$ from the spin-wave Goldstone mode have no anomaly across $T_c$ \cite{CLZhang14}.

\begin{figure}[t] \includegraphics[scale=.35]{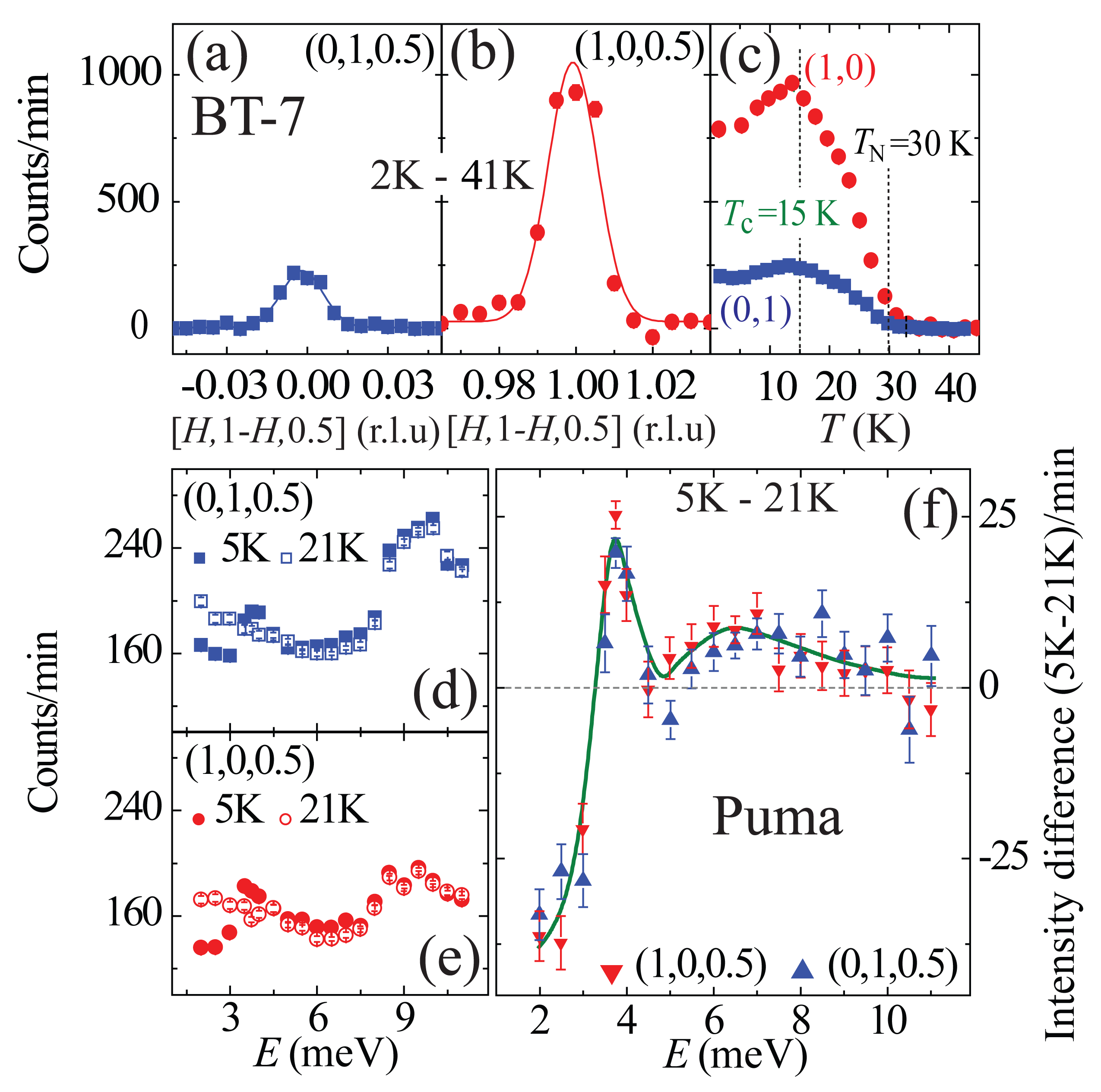}
\caption{(Color online) 
Temperature differences of the elastic scattering below (2 K) and above (41 K) the N$\rm \acute{e}$el temperature ($T_N$ = 33 K)
at reciprocal positions (a) (0,1,0.5) and (b) (1,0,0.5) in a uniaxial strained NaFe$_{0.985}$Co$_{0.015}$As single crystal. 
The uniaxial strain along the $b$ axis is $\sim$ 10 MPa and data was 
collected on BT-7.  By comparing the area ratio of the two peaks, we estimate
that $\sim$ 58$\%\ [\approx (I(1,0,0.5)-I(0,1,0.5))/(I(1,0,0.5)+I(0,1,0.5))]$ of the crystal is detwinned. The solid lines are Gaussian fits to the data. 
Similar detwinning ratio is also obtained at PUMA.  After releasing the uniaxial strain, we find that
the sample returns to the twinned state.
(c) Temperature dependence of the
magnetic order parameters at ${\bf Q}_1$ and ${\bf Q}_2$ in unixially strained detwinned NaFe$_{0.985}$Co$_{0.015}$As.
(d) and (e) Constant {\bf Q}-scans at ${\bf Q}_1$ and ${\bf Q}_2$ below and above $T_c$, respectively, in the partially 
detwinned sample. The peaks around $\sim$9 mV in the raw data are mostly temperature independent 
background scattering.
(f) Temperature differences between 5 K and 21 K, revealing the
superconductivity-induced resonance at ${\bf Q}_1$ and ${\bf Q}_2$.
The nearly identical resonances at these two vectors suggests that the coupling between spin waves
and superconductivity is weak in Co-doped NaFeAs. The solid line is a guide to the eye.
}
\end{figure}

\begin{center}
\textbf{II. Experimental Results and Theoretical Calculations}
\end{center}

We prepared single crystals of NaFe$_{0.985}$Co$_{0.015}$As by the self-flux method \cite{CLZhang13} and cut a large crystal
 into the rectangular shape along the $[1,0,0]$ and $[0,1,0]$ directions ($16.11\times 8.41\times 1.31$ mm$^3$, $\sim$0.79 g).
From NMR measurements \cite{LMa2014}, we know that the tetragonal-to-orthorhombic structural transition happens around
$T_s\approx 40$ K, above $T_N$ and $T_c$. 
Our neutron scattering experiments were carried out on the PUMA and BT-7 thermal 
triple-axis spectrometers at the MLZ, TU M\"{u}chen, Germany \cite{schi09}, and NIST center for neutron research (NCNR),
Gaithersburg, Maryland \cite{jeff}, respectively.  In both cases, we used vertically and horizontally focused pyrolytic monochromator and analyzer with a
fixed final neutron energy of $E_f=14.7$ meV.
The wave vector ${\bf Q}$ at ($q_x$,$q_y$,$q_z$) in \AA$^{-1}$ is
defined as (\textit{H},\textit{K},\textit{L}) = ($q_xa/2\pi$,$q_ya/2\pi$,$q_zc/2\pi$) reciprocal lattice unit (r.l.u.)
using the orthorhombic unit cell ($a \approx b\approx 5.589$ \AA\ and $c = 6.980$ \AA\ at 3 K).  In this notation,
the AF Bragg peaks occur at the $(1,0,L)$ positions with $L=0.5,1.5,\cdots$ and there are no magnetic peaks at $(0,1,L)$ [Figs. 1(a) and 1(b)] \cite{slli09}.
We have used a detwinning device similar to that of the previous INS work on BaFe$_{2-x}$Ni$_x$As$_2$ \cite{xylu14s}.
The samples are aligned in the $[1,0,0.5]\times [0,1,0.5]$ scattering plane.  In this scattering geometry, we can probe the static AF order and
spin excitations at both ${\bf Q}_1$ and ${\bf Q}_2$, thus allowing a conclusive determination of the detwinning ratio and spin excitation anisotropy
at these wave vectors.  Figure 2(a) and 2(b) shows the temperature differences in 
transverse elastic scans along the
$[H,1-H,0.5]$ and $[H,1-H,0.5]$ directions, respectively, for NaFe$_{0.985}$Co$_{0.015}$As between 2 K and 41 K.  By comparing the scattering intensity at these two
wave vectors, we estimate that the sample is about 58\% detwinned.  Figure 2(c) shows the temperature dependence of the magnetic order parameters.
Consistent with previous data on a twinned sample \cite{CLZhang13}, the uniaxial strain used to detwinn the sample does not seem to alter
$T_N\approx 30$ K and $T_c=15$ K.

\begin{figure}[t] \includegraphics[scale=.35]{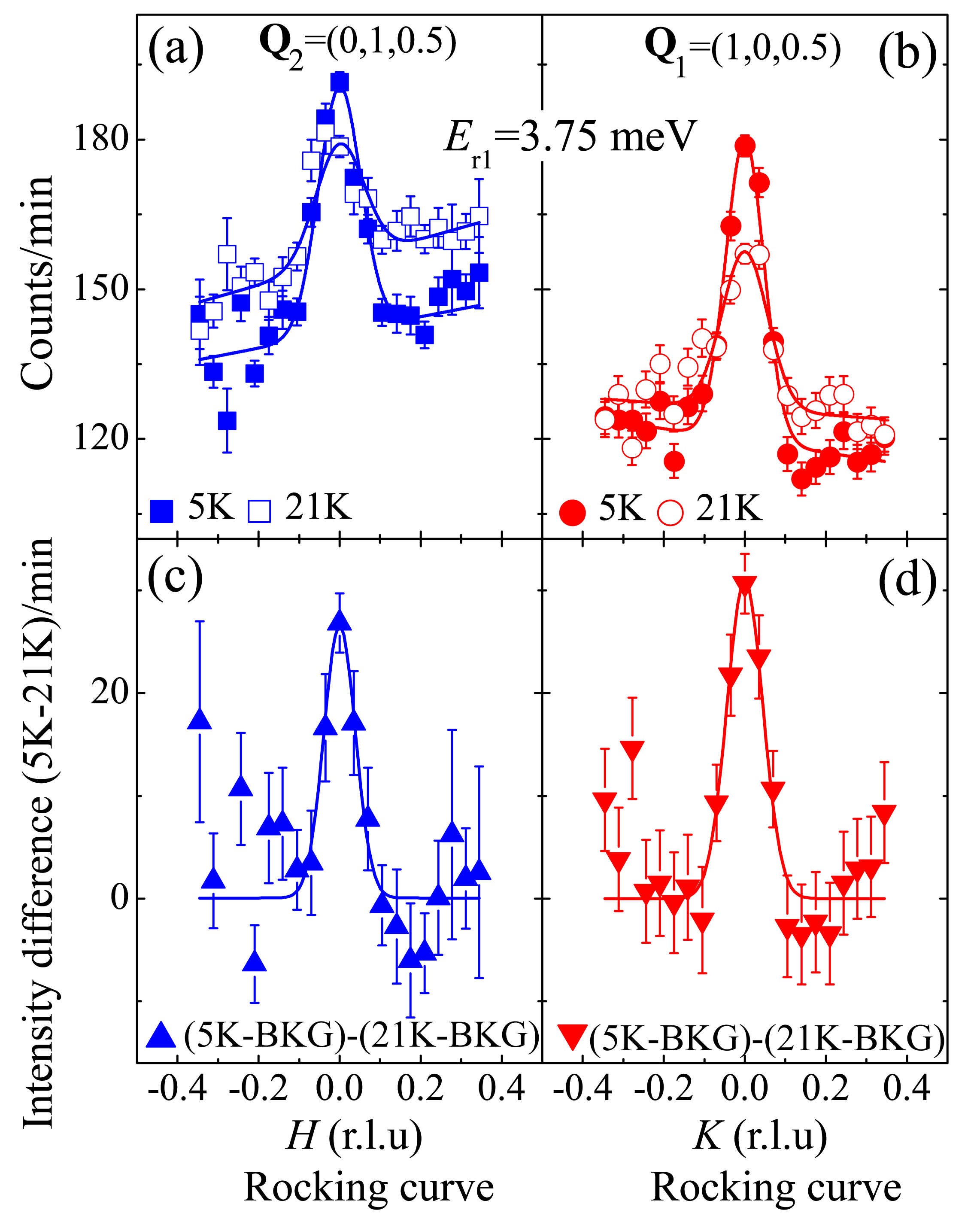}
\caption{(Color online) 
Constant-energy scans at the low-energy resonance $E_{r1}$=3.75 meV for detwinned NaFe$_{0.985}$Co$_{0.015}$As.
(a) and (b) The rocking scans across ${\bf Q}_2=(0,1,0.5)$ and ${\bf Q}_1=(1,0,0.5)$ positions above (21 K) and below (5 K) $T_c$, respectively. The wave vector independent background scattering increases slightly on warming. 
(c) and (d) Corresponding difference between the two temperatures after the background subtraction.
The solid lines are fits to Gaussians on flat linear backgrounds set to zero.}
\end{figure}

In previous INS work on twinned NaFe$_{0.985}$Co$_{0.015}$As, superconductivity induces a dispersive sharp resonance near
$E_{r1}=3.25$ meV and a broad dispersionless mode at $E_{r2}=6$ meV at ${\bf Q}_1=(1,0,0.5)$ and ${\bf Q}_2=(0,1,0.5)$ \cite{CLZhang13}.
To explore what happens in the uniaxial strain detwinned NaFe$_{0.985}$Co$_{0.015}$As,
 we carried out constant-\textbf{Q} scans at wave vectors ${\bf Q}_2$ [Fig. 2(d)] and
${\bf Q}_1$ [Fig. 2(e)] below and above $T_c$.
While it is difficult to see the resonance in the raw data, the temperature differences
between 5 K and 21 K plotted in Fig. 2(f) reveal a sharp peak
at $E_{r1}=3.75$ meV and a broad peak at $E_{r2}=6$ meV, in addition to the negative
scattering below 3 meV due to a spin gap.  These data indicate the presence of a superconductivity-induced sharp resonance
and a broad resonance above a spin gap, similar to the results on twinned samples \cite{CLZhang13}.
Surprising, there are no observable differences for the resonance at ${\bf Q}_1$ and ${\bf Q}_2$, suggesting
that the double resonance is not directly associated with the twinning state of the sample.

\begin{figure}[t] \includegraphics[scale=0.38]{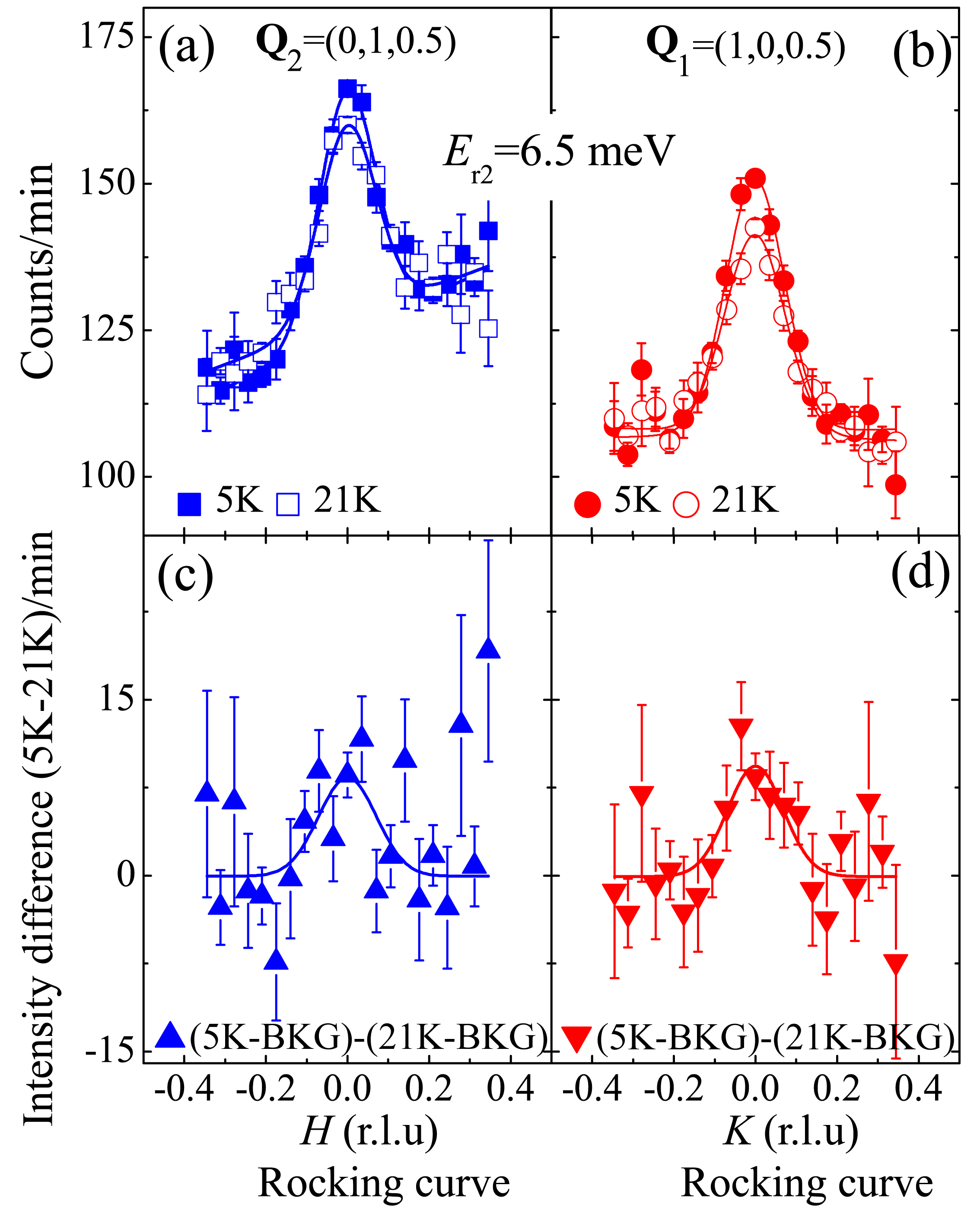}
\caption{(Color online)
Constant-energy scans through the high-energy resonance $E_{r2}$=6.5 meV for detwinned NaFe$_{0.985}$Co$_{0.015}$As.
(a) and (b) The rocking scans across the ${\bf Q}_2=(0,1,0.5)$ and ${\bf Q}_1=(1,0,0.5)$ positions above (21 K) and below (5 K) $T_c$, respectively.
(c) and (d) The corresponding difference between the two temperatures.
The solid lines are fits to Gaussians on zero backgrounds.
}
\end{figure}

To confirm the conclusion of Fig. 2, we carried out constant-energy scans
near  ${\bf Q}_1$ and ${\bf Q}_2$ at $E_{r1}$ and $E_{r2}$
above and below $T_c$.  Figure 3(a) and 3(b) shows transverse rocking curve scans
through ${\bf Q}_2=(0,1,0.5)$ and ${\bf Q}_1=(1,0,0.5)$, respectively,
at the sharp resonance energy $E_{r1}=3.75$ meV above and below $T_c$.
While there is slightly more magnetic scattering at the AF ordering wave vector
${\bf Q}_1=(1,0,0.5)$ compared with that at ${\bf Q}_2=(0,1,0.5)$
in both the normal and superconducting states \cite{xylu14s}, the superconductivity induced intensity changes
shown in Fig. 3(c) and 3(d), defined as the resonance \cite{Eschrig},
at these two wave vectors are indistinguishable within the statistics of our measurement.

Figures 4(a-d) summarize wave vector scans at the energy of
the broad resonance $E_{r2}=6$ meV around
${\bf Q}_2=(0,1,0.5)$ and ${\bf Q}_1=(1,0,0.5)$.
Similar to data at $E_{r1}=3.75$ meV, we find that the superconductivity-induced
intensity gain of the broad resonance is almost indistinguishable at these wave vectors, again
confirming the notion that the resonance is not sensitive to the twinning state of the system.

In electron doped NaFe$_{1-x}$Co$_x$As, the dominant orbital
character of the electron pockets would be $d_{xy/xz}$ at $(1,0)$ and
$d_{yz/xy}$ at $(0,1)$ in the Brillouin zone
[Fig. 1(c)]  \cite{qqge,Liu_arpes,thirupathaiah}.  The orbital character of the hole pocket
is $d_{xz/yz}$. If the superconducting pairing amplitudes are highly orbital dependent,\emph{i.e.}, $\Delta_{xy}\neq\Delta_{xz/yz}$, the superconducting gap can be anisotropic along the electron pocket and this gap anisotropy gives rise to a splitting of the resonance peak \cite{RYu14}. 
Such an orbital-selective pairing scenario is consistent with both ARPES measurements~\cite{qqge} and INS results in 
twinned samples~\cite{CLZhang13}.

\begin{figure}[t] \includegraphics[scale=0.3]{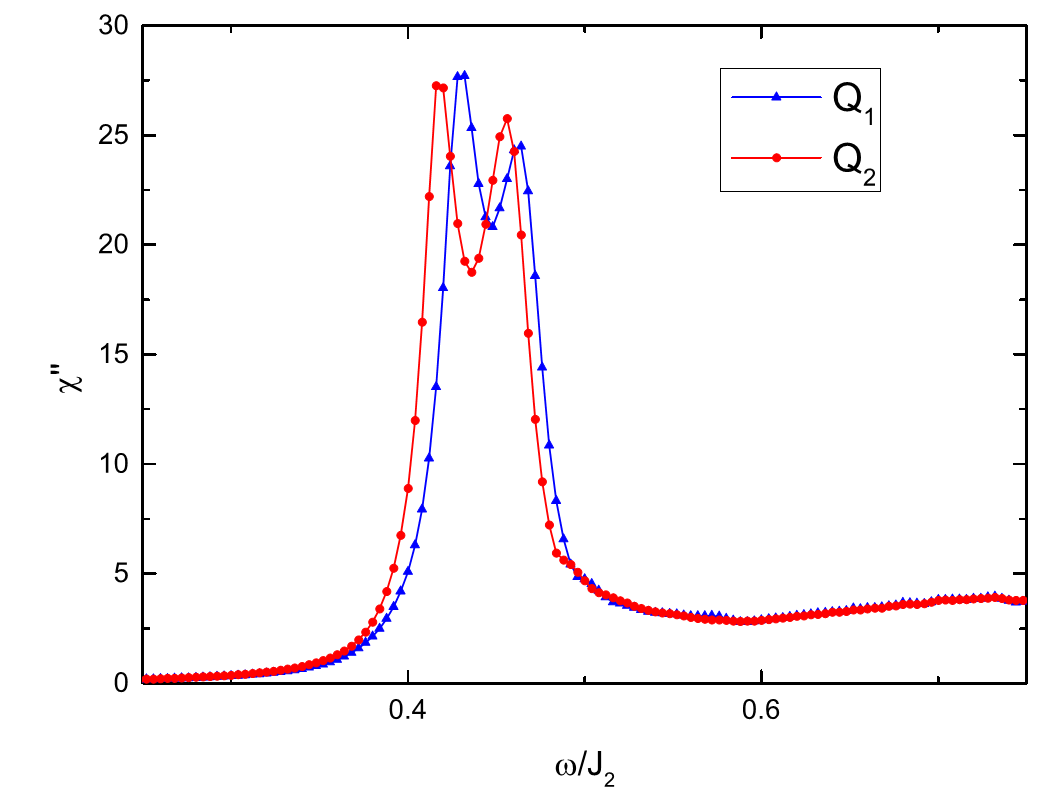}
\caption{
Imaginary part of the dynamic spin susceptibility $\chi^{\prime\prime}$ in the superconducting state of a multiorbital $t-J_1-J_2$ model with a nonzero splitting between the $d_{xz}$ and $d_{yz}$ orbitals, $\epsilon=0.02$ eV. Two resonance peaks are present in each of the ${\bf Q}_1$ and ${\bf Q}_2$ wave vectors. In the calculation, $J_1/J_2=0.1$ is taken such that the pairing amplitudes show strong orbital selectivity. See Ref.~\cite{RYu14} for details of the model and the method.
}
\end{figure}

In the uniaxial strain detwinned sample, the degeneracy of the $d_{xz}$ and $d_{yz}$ orbitals is lifted and, 
correspondingly, the Fermi surface is distorted  
\cite{qqge}. 
To investigate whether the double resonances in the orbital-selective pairing scenario still exist in the presence of a splitting between the $d_{xz}$ and $d_{yz}$ orbitals, we calculated the imaginary part of the spin susceptibility $\chi^{\prime\prime}$ in the superconducting state from a multiorbital $t-J_1-J_2$ model with an orbital splitting term $\epsilon=n_{xz}-n_{yz}$ \cite{RYu14}. Our result for a strong orbital selectivity is presented in Fig. 5. We find two resonance peaks at each of the wave vectors
 ${\bf Q}_1$ and ${\bf Q}_2$ for a nonzero splitting $\epsilon$.  The intensities of the counterpart peaks at ${\bf Q}_1$ and ${\bf Q}_2$ are comparable. At each resonance peak, there is a relative shift of the resonance energy between the ${\bf Q}_1$ and ${\bf Q}_2$ resonances. This shift is proportional to the splitting $\epsilon$.
The calculated double-resonances feature at both ${\bf Q}_1$ and ${\bf Q}_2$ is qualitatively consistent with the experimental observation in the detwinned sample. 
The experiment can not resolve a relative shift of the resonance energy. This could be because either the splitting $\epsilon$ is small in the detwinned underdoped compound, or the coupling between the superconductivity and the splitting $\epsilon$ is rather weak. Further comparison between theory and experiments is needed to fully settle the issue.

\begin{center}
\textbf{III. Conclusion}
\end{center}

In conclusion, our INS experiments on partially detwinned NaFe$_{0.985}$Co$_{0.015}$As reveal the presence of two resonances at
each of the
 wave vectors ${\bf Q}_{\rm AF}={\bf Q}_1=(1,0)$ and ${\bf Q}_2=(0,1)$.
This is different from the scenario where the two resonances are due to
the coexisting AF order with superconductivity \cite{knolle11,WCLv14}.
Instead, the data are qualitatively consistent with 
the proposal that the double resonances originate from an
orbital dependence of the superconducting pairing.
Our results provide further evidence that
orbital selectivity
plays an important role in understanding 
not only the  normal state but also 
the superconducting pairing of the multiorbital electrons in the iron pnictides.

\begin{center}
\textbf{IV. Acknowledgements}
\end{center}

We thank Z. C. Sims for his help in single crystal growth efforts.
The single crystal growth and neutron scattering
work at Rice is supported by the U.S. DOE, BES under contract no. DE-SC0012311 (P.D.).
Part of the work is also supported by the Robert A. Welch foundation grant numbers C-1893 (P.D.) and C-1411 (Q.S.).
Q.S. is also supported by US NSF DMR-1309531.
R.Y. was supported by the NSFC grant No. 11374361, and the Fundamental Research Funds for the Central Universities and the Research Funds of Renmin University of China.

\end{document}